\documentclass[11pt, sf]{iopart}

\usepackage[english]{babel}
\usepackage[T1]{fontenc}
\usepackage{amssymb,amstext}
\usepackage{graphicx}
\usepackage{geometry}%
\usepackage{url}
\usepackage{texdraw}
\usepackage{hyperref}
\usepackage{xspace}
\usepackage[labelfont={sf,bf,footnotesize}, textfont={footnotesize}]{caption}
\usepackage{placeins}

\usepackage{psfrag}
\usepackage{rotating}
\usepackage{color}
	 \definecolor{darkred}{rgb}{0.75,0,0}
	 \definecolor{darkgreen}{rgb}{0,0.5,0}
	 \definecolor{darkblue}{rgb}{0,0,0.75}

\newcount\hour \newcount\minute
\hour=\time \divide \hour by 60 \minute=\time \count99=\hour
\multiply \count99 by -60 \advance\minute by \count99 \hour=\time
\divide \hour by 60

\date{\today}

\newcommand{\ie}{i.\,e.\,\,}
\newcommand{\eg}{e.\,g.\,\,}
\newcommand{\del}{\partial}
\newcommand{\Del}[2]{\frac{\partial #1}{\partial #2}}

\newcommand{\EXP}[1]{\exp\left\{ #1 \right\}}
\newcommand{\eq}[1]{(\ref{eq:#1})}
\newcommand{\fig}[1]{\ref{fig:#1}}
\newcommand{\at}[1]{\big|_{_{#1}}}
\newcommand{\At}[1]{\Big|_{_{#1}}}

\begin{document}

\title{Fixation times in evolutionary games under weak selection}
\author{Philipp M. Altrock and Arne Traulsen}
\address{Max Planck Institute for Evolutionary Biology, August-Thienemann-Str.\,2, 
24306 Pl\"on, Germany\\ {\it altrock@evolbio.mpg.de, traulsen@evolbio.mpg.de}}
\date{\today, \number\hour:\number\minute}

\begin{abstract}

In evolutionary game dynamics, reproductive success increases with the performance in
an evolutionary game.
If strategy $A$ performs better than strategy $B$, strategy $A$ will spread in the population.
Under stochastic dynamics, a single mutant will sooner or later take over the entire population or
go extinct.
We analyze the mean exit times (or average fixation times) associated with this process.
We show analytically that these times
depend on the payoff matrix of the game in an amazingly simple way under weak selection, \ie strong stochasticity:
The payoff difference $\Delta \pi$ is a linear function of the number of $A$ individuals $i$,
$\Delta \pi = u \, i + v$.
The unconditional mean exit time depends only on the constant term $v$.
Given that a single $A$ mutant takes over
the population, the corresponding conditional mean exit time depends only on the density dependent term $u$.
We demonstrate this finding for two commonly applied microscopic evolutionary processes.

\end{abstract}

\maketitle

\section{Introduction}

Systems in which successful strategies spread by imitation or genetic reproduction can be described by evolutionary game theory. Such models are routinely analyzed in evolutionary biology, sociology, anthropology and economics. Recently, the application of methods from statistical physics to these systems has lead to many important insights \cite{szabo:2007aa,berg:1998aa,szabo:2002te,santos:2005pm,hauert:2005mm}.

Traditionally, the dynamics is described by the replicator equations, where the growth rate of a strategy is associated with its relative success compared with the population average \cite{taylor:1978wv,hofbauer:1998mm}. 

In the past years, research has focused on stochastic evolutionary game dynamics in finite populations
\cite{nowak:2004pw,taylor:2004wv,traulsen:2005hp,wild:2004aa,imhof:2005oz,traulsen:2006ab,fudenberg:2006fu,reichenbach:2006aa,perc:2006aa,perc:2006bb,perc:2007bb,claussen:2007aa,cremer:2008aa,Szolnoki:2008aa,claussen:2008aa}.
In this context, a connection to the weak selection limit of population genetics has been established
\cite{nowak:2004pw}. Weak selection means that the payoff differences based on different strategic behavior in interactions represent only a small correction to otherwise random dynamics, similar to high temperature expansions in physics. 
Weak selection is considered as a relevant limit in biology, as most evolutionary changes are driven by small fitness differences \cite{ohta:2002aa}. Moreover, it allows analytical approximations that are often impossible when selective differences in payoffs are large \cite{nowak:2004pw,ohtsuki:2006na,ohtsuki:2007pr}.

Most of the recent work that uses the weak selection approximation has been focusing on the probability that a certain strategy takes over.
The time associated with this process has been calculated \cite{antal:2006aa}, but it received
considerably less attention so far. 
Here, we present the weak selection corrections to the conditional and unconditional mean exit or fixation times in evolutionary $2\times2$ games with $N$ players.

The conditional average time to fixation $t_1^A$ is the expected time a single mutant needs to take over the population, given that such a takeover occurs at all.
The unconditional average time of fixation $t_1$ is the expectation value for the time until the population is homogenous again after the arrival of a single mutant. This is regardless of wether the mutant type takes over the population or becomes extinct.
Equivalently, the average fixation times for such one dimensional random walks can also be interpreted as mean first passage times or mean exit times \cite{kampen:1997xg,redner:2001bo, Gardiner04}.

\begin{figure}
\begin{center}
	\includegraphics[width=0.95\textwidth]{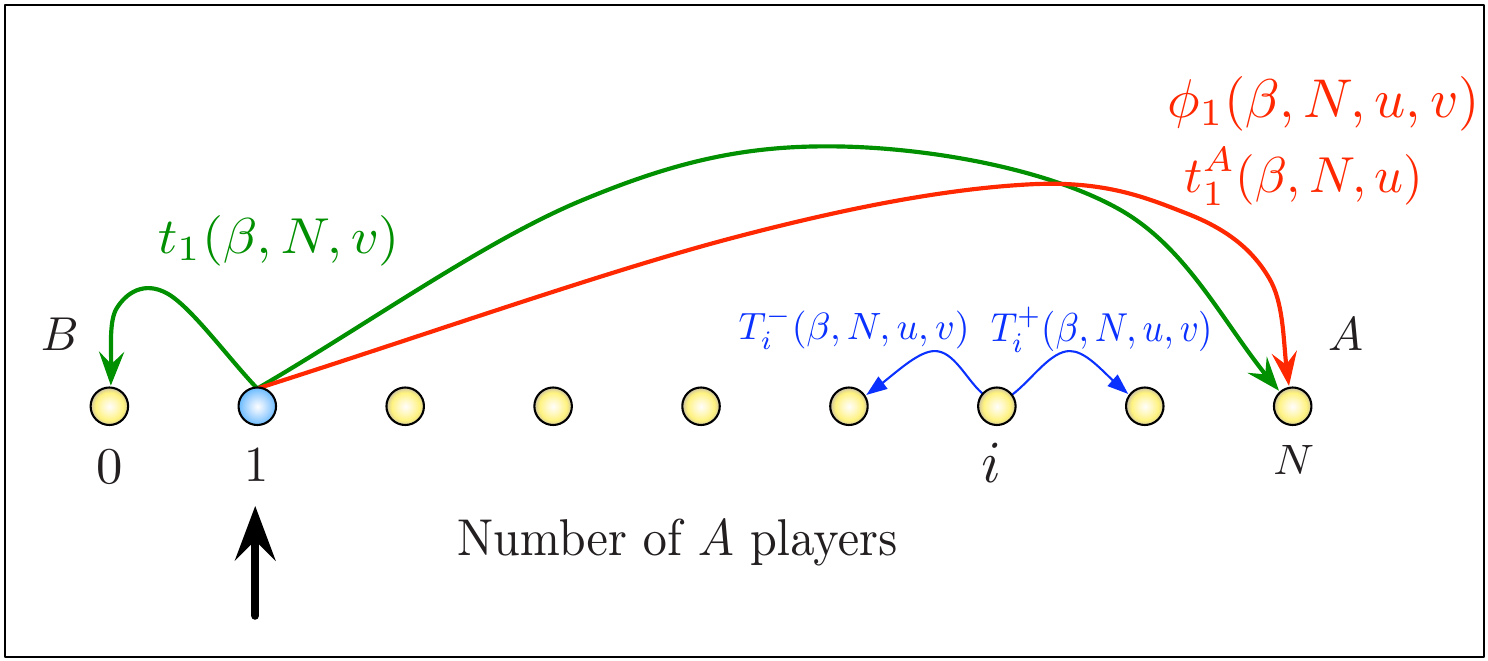}
\caption{
Illustration of the most relevant quantities. 
We are interested in the evolutionary fate of a single $A$ player.
All quantities depend on the intensity of selection $\beta$ and the 
population size $N$.  
The payoff difference between $A$ and $B$ players is
given by $\Delta \pi = u \, i + v$, with $i$ as the number of $A$ players. 
Both the transition probabilities $T^+_i$ and $T^-_i$ 
and the probability that a single $A$ player takes over the population 
$\phi_1$ depend on $ u$ and $v$. 
But for weak selection, $\beta \ll 1$, the conditional time $t_1^A$ during which a single
$A$ player takes over a population of $B$ players only depends on $u$,
whereas the unconditional time $t_1$ until either $A$ or $B$ has taken over the
population only depends on $v$. 
}
\label{fig:FIG1}
\end{center}
\end{figure}

Throughout this paper, we use the payoff matrix
\begin{equation}\label{eq:Pmatrix}
\bordermatrix{
  & A & B \cr
A & a & b \cr
B & c & d \cr}.
\end{equation}
An $A$ player interacting with another $A$ receives $a$. 
If it interacts with $B$, it obtains $b$.
Similarly, $B$ receives $c$ from $A$ and $d$ from other $B$'s. 
Thus, the average payoffs are
\begin{eqnarray}
\pi_A(i) &=& \frac{i-1}{N-1} \, a+  \frac{N-i}{N-1}\, b \label{eq:PiA}\\
\pi_B(i) &=&  \frac{i}{N-1}\,c + \frac{N-i-1}{N-1}\, d\label{eq:PiB}.
\end{eqnarray}
A quantity that is of particular interest is the difference between the average payoffs,
\begin{equation}\label{eq:Delta}
\Delta \pi(i) = \pi_A(i) -\pi_B(i) = u \, i  + v,
\end{equation}
where
\begin{eqnarray}
	u&=&\frac{a+d-(b+c)}{N-1},\label{eq:U} \\ 
	v&=&\frac{N(b-d)-(a-d)}{N-1}.\label{eq:V}
\end{eqnarray}
We show that under weak selection, the conditional time ($t_1^A$) during which a single
mutant takes over the whole population depends only on $u$ (and, of course, on the population size).
The unconditional time ($t_1$) during which the mutant either takes over the population or reaches extinction
depends only on $v$ (and the population size).
See Figure \fig{FIG1} for an illustration of the relevant quantities. 

Our manuscript is organized as follows: In Section \ref{Fermi}, we introduce a particular evolutionary process for our analysis. Although our results are valid for a broader class of processes, we only present the full calculation for this evolutionary process. 
In Section \ref{General}, we recall the general form of fixation probabilities and times.
We discuss neutral selection in Section \ref{Neutral} as a prerequisite to the weak selection expansion,
which we explore in Section \ref{Weak}. 
In Section \ref{Moran} we address the frequency dependent Moran process to underline the generality of our findings.
The consequences of our analytical results are discussed in Section \ref{discussion}.

\section{Fermi process}\label{Fermi}
 
In a finite population of size $N$ with two possible strategies $A$ and $B$, the state of the system is characterized by the number of type $A$ individuals $i$. In general, the dynamics is stochastic.
In each time step, a randomly chosen individual evaluates its sucess. 
It compares this payoff with a second, randomly chosen individual. 
If this second individual has a higher payoff, the first one switches strategies with probability $p>\frac{1}{2}$. Otherwise, it switches with $p<\frac{1}{2}$. 
We assume that
the switching probability is given by the Fermi distribution. 
Its shape is controlled by the intensity of selection $\beta$, which can be interpreted as an inverse temperature,
\begin{eqnarray}\label{eq:Fermi01}
	p_i^{\pm} = \frac{1}{1+e^{\mp\beta(\pi_A(i)-\pi_B(i))}}=\frac{1}{1+e^{\mp\beta\Delta\pi(i)}}.
\end{eqnarray}
In  previous work \cite{blume:1993jf,szabo:1998wv, pacheco:2006pb}, there is a different strategy update procedure.
The first individual switches to the second's strategy with probability $p_i^{\pm}$.
The second individual can also switch to the first individual's strategy with probability $1-p_i^{\pm}$.
This yields a factor $2$ in the transition probabilities (and, as we will become clear later, a factor $\frac{1}{2}$ in the fixation times).
This process also has a proper strong selection limit, \ie it is possible to examine $\beta\to\infty$. In this latter case we have $p^{\pm}_i \to \Theta(\Delta\pi(i))$, where $\Theta(x)$ is the step function.

The population size is constant in time, in each time step the state of the system can at most change by one, \ie from $i$ to $i-1$ or to $i+1$.
The transition probabilities $T_i^\pm$ to move from $i$ to $i\pm1$ are
\begin{eqnarray}\label{eq:TransProb}
	T^{\pm}_{i} = \frac{i}{N}\frac{N-i}{N}\,p_i^{\pm}.
\end{eqnarray}
The probability to stay in the current state is $1- T^{+}_{i}-T^{-}_{i} $.
An important measure of where the system is more likely to move is their ratio, 
\begin{eqnarray}\label{eq:probratio}
	\gamma_{i}=\frac{T_{i}^{-}}{T_{i}^{+}}=\e^{-\beta\,\Delta\pi(i)}. 
\end{eqnarray}
This is a quantity that describes the tendency to move from the state $i$ to $i\mp1$, depending on whether $\gamma_i\gtrless1$. Of course, $T_i^+>0$ is required, which follows from $\beta<\infty$. The $T_i^{\pm}$ and thus the
$\gamma_i$ are invariant under adding a value to each of the payoffs given in \eq{Pmatrix}, whereas multiplying the payoff matrix with a factor $\lambda$ results in a change in the intensity of selection $\tilde\beta=\beta\,\lambda$.

Let us now focus on weak selection, $\beta\ll1$.
In this case we have 
\begin{eqnarray}\label{eq:Fermi02}
	p_i^{\pm} \approx \frac{1}{2} \pm \frac{\beta}{4}\Delta\pi(i).
\end{eqnarray}
Weak selection corresponds to high temperature in Fermi statistics. 
A Taylor expansion of the $\gamma_i$ up to first order in $\beta$
 yields $\gamma_i \approx 1-\beta\,\Delta\pi(i)$. 
 In this case, the probability to move from $i$ to $i+ 1$ is very similar to the probability to move from $i$ to $i- 1$. 
 Weak selection links the Fermi process to a variety of birth death processes, cf.\ \cite{nowak:2004pw,traulsen:2007cc}.

\section{Fixation probabilities and fixation times}
\label{General}

From equation \eq{TransProb} it follows that the two pure states all $A$ or all $B$ are absorbing,
$T_0^{\pm} = T_N^{\pm} = 0$. 
In a finite population, we can calculate the probability $\phi_i$ that the system will fixate to the pure state all $A$, starting with the mixed state $i$. Obviously, we have $\phi_{0}= 0$ and $\phi_{N}=1$. For $0<i<N$,
there is a balance equation for the fixation probabilities, $\phi_{i}=T_{i}^{-}\phi_{i-1}+(1-T_{i}^{+}-T_{i}^{-})\phi_{i}+T_{i}^{+}\phi_{i+1}$. This recursion leads to an expression for the fixation probabilities in terms of the $\gamma_i$ \cite{karlin:1975xg,traulsen:2009bb,nowak:2006bo}, 
\begin{eqnarray}\label{eq:Fixprob01}
	\phi_{i}=\frac{1+\sum_{k=1}^{i-1}\prod_{l=1}^{k}\gamma_{l}}{1+\sum_{k=1}^{N-1}\prod_{l=1}^{k}\gamma_{l}},
\end{eqnarray}
which is valid for any birth death process. 

For the Fermi process, the exact equation \eq{probratio} simplifies matters in an elegant way because the products in equation \eq{Fixprob01} can be solved,
\begin{eqnarray}\label{eq:gamma02}
	\prod\limits_{l=1}^{k}\gamma_{l}=\exp\left\{-\beta\,\sum\limits_{l=1}^{k}\Delta\pi(l) \right\}=\exp\left\{-\beta\,\left[ k^2\frac{u}{2}+k \left(\frac{u}{2}+v \right) \right] \right\}.
\end{eqnarray}
Hence, equation \eq{Fixprob01} simplifies to
\begin{eqnarray}\label{eq:Fixprob02}
	\phi_i=\frac{1+\sum_{k=1}^{i-1}\EXP{-\beta \left[k^2\frac{u}{2}+k(\frac{u}{2}+v) \right]}}{1+\sum_{k=1}^{N-1}\EXP{-\beta \left[k^2\frac{u}{2}+k(\frac{u}{2}+v) \right]}}.
\end{eqnarray}
For large $N$, the sums in equation \eq{Fixprob02} can be approximated by integrals, which yields a closed expression for the probabilities $\phi_i$  \cite{traulsen:2007cc,traulsen:2006bb}.

General expressions for the unconditional and conditional mean exit times or average times of fixation, $t_1$ and $t_1^A$, are well known, especially for 
simple, translational invariant random walks \cite{antal:2006aa,kampen:1997xg,Fisher1988aa}. 
A complete derivation for the
average times of fixation in finite systems 
without translational invariance
can be found in \cite{antal:2006aa,traulsen:2009bb,goel:1974aa}.
 
In the following, we will focus on the fixation of a single $A$ mutant in a population of $B$.
Accordingly, the unconditional and conditional fixation times read
\begin{eqnarray}\label{eq:Ufixtime1}
	t_1=\phi_1\sum\limits_{k=1}^{N-1}\sum\limits_{l=1}^k\frac{1}{T_l^+}\prod\limits_{m=l+1}^{k}\gamma_m,
\end{eqnarray}
and
\begin{eqnarray}\label{eq:Cfixtime1}
	t_1^A =  \sum\limits_{k=1}^{N-1}\sum\limits_{l=1}^k\frac{\phi_l}{T_l^+}\prod\limits_{m=l+1}^{k}\gamma_m,
\end{eqnarray}
respectively. 
Time is measured in elementary time steps here.
Thus, in each time step one reproductive event occurs. 
In biological contexts, it is often more convenient to measure
time in generations, such that  each individual
reproduces once per generation on average. 
Time in generations is obtained
by dividing the number of time steps by the population size $N$. 
It is well known that the variance of the exit times under weak selection can be large \cite{goel:1974aa}, which has important biomedical implications \cite{dingli:2007aa}.
Nonetheless, here we concentrate on the expectation values and do not address the distribution of the exit times.

\section{Neutral selection}
\label{Neutral}

An important reference case is neutral selection, which results from vanishing selection intensity $\beta=0$ \cite{kimura:1968aa}. 
Neutral selection is a very general limit, which is typically not affected by the details of the evolutionary process.
For neutral selection we have $\gamma_i=1$, that is $T_i^+=T_i^-$ in any state $i$. However, we still have $T_i^{\pm}\neq T_j^{\pm}$ for $i\neq j$, although the system is symmetric, $T^{\pm}_{i}=T_{N-i}^{\pm}$. This is a difference to the simple random walk in one dimension, which is invariant with respect to translation \cite{Gardiner04}.

For the Fermi process, the neutral transition probabilities are
\begin{eqnarray}\label{eq:TprobN}
	T_i^{\pm}\At{\beta=0}=\frac{1}{2}\frac{i}{N}\frac{N-i}{N}.
\end{eqnarray}
We have $T_i^{+} = T_i^{-}$, which leads to $\gamma_i=1$. 
From equation \eq{Fixprob01}, it is thus clear that 
the probability of fixation to $A$ is given by the initial abundance of $A$, 
\begin{eqnarray}\label{eq:FixprobN}
	\phi_i\At{\beta=0}=\frac{i}{N}.
\end{eqnarray}
For the neutral unconditional time of fixation $t_1$ we get
\begin{eqnarray}\label{eq:t1N01}
	t_1\At{\beta=0}=\frac{1}{N}\,\sum\limits_{k=1}^{N-1}\sum\limits_{l=1}^k\frac{2\,N^2}{l(N-l)}=2\,N\,H_{N-1}.
\end{eqnarray}
Details for this calculation can be found in \ref{app:general}. We introduced the shorthand notation for the harmonic numbers $H_{N-1}=\sum_{l=1}^{N-1} \frac{1}{l}$, which diverge logarithmically with $N$. In the same way we can solve 
\begin{eqnarray}\label{eq:t1AN01}
	t_1^A\at{\beta=0} &= \sum\limits_{k=1}^{N-1}\sum\limits_{l=1}^k\frac{l}{N}\frac{2\,N^2}{l(N-l)} = 2\,N(N-1).
\end{eqnarray}
For neutral selection, the conditional average time of fixation of a single mutant diverges quadratically with the system size.

\section{Weak Selection}
\label{Weak}

In this section we will calculate the linear corrections of the mean exit times or fixation times $t_1$,  and $t_1^A$ under weak selection, $\beta\ll1$.
Of course, all weak selection approximations are valid only if the term linear in $\beta$ is small compared to the constant term.

The fixation probabilities for small $\beta$ are
\begin{eqnarray}\label{eq:FixprobL01}
	\phi_i\approx\frac{i}{N}+\frac{i}{N}(N-i)\frac{(N+i)u+3v}{6}\beta,
\end{eqnarray}
which has been derived for a variety of evolutionary processes before \cite{nowak:2004pw,traulsen:2005hp,imhof:2005oz,traulsen:2009bb,nowak:2006bo,lessard:2007aa}.

Next, we address the weak selection approximation of the fixation times. 
The expectation value of the unconditional fixation time of a single $A$ mutant in a population of $B$ is in general given by the exact equation \eq{Ufixtime1}.
With the the transition and fixation probabilities of the Fermi process,
the unconditional fixation time of absorption at any boundary simplifies to
\begin{eqnarray}\label{eq:weak01}
	t_1= \phi_1\sum\limits_{k=1}^{N-1}\sum\limits_{l=1}^k \frac{ N^2}{l(N-l)} &&\left(1+e^{-\beta(u\,l+v)}\right)\nonumber\\&&\times  \EXP{-\beta\sum\limits_{m=l+1}^{k}\Delta\pi(m)}.
\end{eqnarray}
The weak selection approximation
takes the remarkably simple form (see \ref{app:weakA} for details)
\begin{eqnarray}\label{eq:weakUncond01}
	t_1 \approx  2\,NH_{N-1} + v\,N\left( N-1-H_{N-1} \right)\,\beta,
\end{eqnarray}
with $v$ given in (\ref{eq:V}). Thus, $t_1$ depends only on the constant term of the payoff difference. For large $N$, this yields $v\approx b-d$. That is, for large populations under weak selection the linear correction of the average 
fixation time only depends on the advantage (or disadvantage) of the $A$ mutants in the resident population. 
For $b>d$, invasion of $A$ mutants is likely and slows down the time until the population is homogeneous again. 
For $d>b$, it is difficult for $A$ to invade a $B$ population and extinction of the mutants is faster than in the neutral case. Note that the payoff entries $a$ and $c$ have no influence on the unconditional fixation time under weak selection corrections.
Since fixation is unlikely for weak selection (the probability of fixation of a single $A$ mutant is approximately $N^{-1}$),
the unconditional fixation time is dominated by the fixation to $B$. In this case, it is enough to discuss the invasion of $A$ mutants.

Next, we address the average time to fixation given that the $A$ mutant takes over the population. 
With the general result \eq{Cfixtime1} the Fermi processes conditional fixation time to all $A$ reads
\begin{eqnarray}\label{eq:weak03a}
	t_1^A = \sum\limits_{k=1}^{N-1}\sum\limits_{l=1}^k\phi_l \frac{ N^2}{l(N-l)} &&\left(1+e^{-\beta(u\,l+v)}\right)\nonumber\\&&\times   \EXP{-\beta\sum_{m=1+l}^{k}\Delta\pi(m)}.
\end{eqnarray}
Its linear approximation turns out to be dependent on the payoffs in a very 
simple way as well,
\begin{eqnarray}\label{eq:weakCond01}
	t^A_1 \approx 2\,N(N-1) - u\,N(N-1)\frac{N^2+N-6}{18}\,\beta,
\end{eqnarray} 
with $u(N-1)= a-b-c+d$. The detailed calculation can be found in \ref{app:weakA}.
Since during the fixation process all payoffs are of importance, it is obvious that they all enter
here. For example, when it is easy to invade because few mutants have an advantage ($b>d$), 
but difficult to reach fixation because mutants are disadvantageous once they are frequent ($c>a$), 
we have $u<0$ and
the conditional time to fixation is larger than neutral. 
In the last section, we discuss special classes of games to show that,
under weak selection,
the conditional mean exit times of fixation (or absorption)  
do not always follow the intuition based on the payoff matrix \eq{Pmatrix}.

\section{Frequency dependent Moran process}
\label{Moran}

In this section we address the generality of the previous findings discussing an alternative evolutionary process.
The first model that connects payoffs from a $2\times2$ game to reproductive fitness using a weak selection approach in finite populations is the frequency dependent Moran process \cite{nowak:2004pw,taylor:2004wv}. In this process, an individual is
chosen for reproduction with probability proportional to its fitness $f(i)$.
The offspring replaces a randomly chosen individual. The average payoffs \eq{PiA} and \eq{PiB} are mapped to the fitness such that 
$f_A(i) = 1 - \beta + \beta\,\pi_A(i)$
and
$f_B(i) = 1 - \beta + \beta\,\pi_B(i)$, where the selcetion intensity $\beta>0$ is so small that $f_A(i)>0$ and $f_B(i)>0$.
The transition probabilities of the standard Moran process read
\begin{eqnarray}
	T_{i}^{+} &=& \frac{i f_{A}(i)}{i f_{A}(i)+(N-i) f_{B}(i)}\frac{N-i}{N},\label{eq:MoranT01}\\
	T_{i}^{-} &=& \frac{(N-i) f_{B}(i)}{i f_{A}(i)+(N-i) f_{B}(i)}\frac{i}{N}\label{eq:MoranT02}.
\end{eqnarray}
Although these transition probabilities are different from those of the Fermi process, 
they also yield $\gamma_{i}\approx1-\Delta\pi(i)$
and $\prod_{m=l+1}^{k}\gamma_{m}\approx1-\beta\sum_{m=l+1}^{k}\Delta\pi(m)$ 
for weak selection, $\beta\ll1$.
Thus, the weak selection approximations of the fixation probabilities $\phi_{l}$ of the Moran process 
and the Fermi process are identical, see equation \eq{FixprobL01}.  
But the weak selection approximation of the transition probabilities are not identical, which leads consequently 
to different mean exit times. 
Nevertheless, the results have the same, remarkably simple connection to the payoff matrix \eq{Pmatrix}. 
The mean exit times or fixation times of the frequency dependent Moran process are
\begin{eqnarray}
	t_{1}&\approx&NH_{N-1}+v\,\frac{N}{2}\left( N+1-2H_{N} \right)\,\beta,\label{eq:Morantime02}\\
	t_{1}^{A}&\approx&N(N-1)-u\,\frac{N^{2}(N^{2}-3N+2)}{36}\,\beta.\label{eq:Morantime03}
\end{eqnarray}
Qualitatively, the dependence on the payoff matrix via $u$ and $v$ is the same as for the Fermi process.
Their calculation is analogous to the findings of the previous section, details can be found in \ref{app:weakA}. Note that, comparing with the Fermi process, there is a factor of $2$ missing in the neutral terms. However, this can be avoided by rescaling the transition probabilities, without changing the properties of the different processes.

\section{Discussion}\label{discussion}

Finally, let us discuss the implications of our results for general $2\times2$ games. 
While we concentrate on the Fermi process here, the discussion is equally valid for
the frequency dependent Moran process.
An important question is whether the linear correction for weak selection is compatible
with the general features of the game and the known asymptotic behavior for large $N$ of the mean exit or fixation times derived
by Antal and Scheuring \cite{antal:2006aa}. 
Clearly, this depends on the payoff matrix of the $2\times2$ game, 
\begin{equation}\label{eq:PmatrixD}
\bordermatrix{
  & A & B \cr
A & a & b \cr
B & c & d \cr},
\end{equation}
as the payoffs enter the first exit times of absorption linearly.
To analyze the difference to the neutral case we consider the rescaled average times of fixation, $\tau_1(\beta)=t_1(\beta)/t_1(0)$ and $\tau_1^A(\beta)=t_1^A(\beta)/t_1^A(0)$. The rescaled unconditional fixation time reads
\begin{eqnarray}\label{eq:weakUncond01R}
	\tau_1 \approx  1 + \frac{1}{2}\frac{N(b-d)-a+d}{N-1}\,\left( \frac{N-1}{H_{N-1}}-1 \right)\,\beta.
\end{eqnarray}
Accordingly, the rescaled conditional fixation time for absorption at all $A$ is
\begin{eqnarray}\label{eq:weakCond01R}
	\tau^A_1 \approx 1 - \frac{a-b-c+d}{N-1}\,\frac{N^2+N-6}{36}\,\beta.
\end{eqnarray}
Note that for population sizes $N>2$ and sufficiently small $\beta$, we always have $t_1(0)<t_1^A(0)$. 
In other words, the average time until the $A$ individual has reached fixation or gone extinct is
smaller than the conditional average time until the $A$ individual has reached fixation. 
For $\beta\to\infty$, the process follows deterministically the intensity of selection and thus both fixation times
may coincide, $t_1(\beta \to \infty) \approx t_1^A(\beta \to \infty)$. This ordering of the fixation times is blurred by our rescaling, as we focus only on the change relative to the neutral case.

In the following, we discuss these two expressions for the three generic types of $2 \times2$ games, 
namely dominance of $A$ ($a>c$ and $b>d$), coexistence of $A$ and $B$ ($a<b$ and $c>d$) 
and a coordination game ($a>c$ and $b<d$).

\subsection{Dominance of $A$.}
Consider a game where strategy $A$ is always dominant, \ie it obtains a larger payoff than $B$, regardless
of the fraction of $A$ in the population. This is the case for $a>c$ and $b>d$.
One special case is the Prisoner's Dilemma with $b>d>a>c$.
The interesting feature of this game is that the social optimum $d$ is not the Nash equilibrium, which is $a$.
For neutral selection, a single $A$ individual goes extinct with probability $1-N^{-1}$. 
Thus, the unconditional
fixation time $\tau_1$ is dominated by the extinction of $A$. 
Since strategy $A$ is favored by selection, increasing the intensity of selection decreases the probability of 
the extinction of $A$. Since fixation takes at least $N-1$ time steps, $\tau_1$ increases with increasing intensity
of selection $\beta$. 
For large $N$, this is obvious from our equation \eq{weakUncond01R}, because in this case the quantity $N(b-d)-a+d$
is positive.
However, once extinction of $A$ becomes unlikely, increasing $\beta$ further will lead to a
decrease of $\tau_1$. 

The discussion of the conditional fixation time $\tau_1^A$ is not as straightforward, 
because the sign of $a-b-c+d$ can be positive or negative. The sign of this quantity 
is also decisive for the evolutionary dynamics in other contexts, see e.g.\ \cite{taylor:2007bb}. 
When the advantage of an $A$ individual is initially large and decreases with the abundance of $A$
($a-c> b-d>0$), then the sign of $a-b-c+d$ is positive and $\tau_1^A$ decreases with increasing intensity
of selection. 
But when the advantage of strategy $A$ decreases with the number of $A$ individuals
($b-d>a-c>0$), then $\tau_1^A$ increases with increasing intensity of selection. 
However, this apparently counterintuitive phenomenon (after all, $A$ dominates $B$)
can only be observed for weak selection. For strong selection, $\tau_1^A$ decreases again.
These results are compatible with the observation that the conditional fixation time scales as 
$N \ln N$ for large $N$ \cite{antal:2006aa}. 
 In Figure \fig{FIG2} (a) we show a numerical example for the rescaled average times.
 We include averages from numerical simulations of the evolutionary process, 
 our linear approximation as well as the exact result that can be obtained from dividing equation \eq{Ufixtime1}
 by \eq{t1N01} and equation \eq{Cfixtime1}
 by  \eq{t1AN01}, respectively. 
The payoff matrix is chosen such that $a+d>b+c$, which means that with increasing intensity of selection
$\tau_{1}^{A}$ decreases and $\tau_{1}$ increases. 

{\subsection{Coexistence of $A$ and $B$.}
As a second class, we consider games in which $B$ is the best reply to $A$ ($c>a$), but 
$A$ is the best reply to $B$ ($b>d$). 
Important examples for such games are the Hawk-Dove game \cite{maynard-smith:1973to} or the Snowdrift game \cite{doebeli:2005aa}. For infinite populations, the replicator dynamics predicts a stable coexistence of $A$ and $B$. 
In finite populations, the system typically fluctuates around that point until eventually, fluctuations lead to absorption in one the boundaries \cite{Claussen:2005eh,chalub:2006cc}. 
Consequently, the conditional fixation times increase exponentially with the population size \cite{antal:2006aa}.  
Since $a-b-c+d$ is negative, we also have an increase of $\tau_1^A$ with the selection intensity for weak selection. 
Further, $N(b-d)-a+d$ is positive in large populations, such that also $\tau_1$ increases with the selection intensity.
Figure \fig{FIG2} (b) shows that the divergence of the exact results is faster than the linear approximation even for weak selection.

\subsection{Coordination games.}
Finally, let us discuss coordination games in which $a>c$ and $b<d$. In these games, $A$ is the best reply to $A$ and $B$ is the best reply to $B$. The replicator equation of such systems exhibits a bistability: If the fraction of $A$ individuals is sufficiently high in the beginning, the $A$ individuals will reach fixation. Otherwise, $B$ individuals will take over the system. 
The stronger the intensity of selection, the less likely it is that a single $A$ individual can take over a $B$ population. 
Consequently, $\tau_1$ should decrease with $\beta$. This also follows from our weak selection approximation: 
In large populations, $N(b-d)-a+d$ is negative and thus $\tau_1$ decreases with the intensity of selection, see equation \eq{weakUncond01R}. 
Perhaps less intuitive, also $\tau_1^A$ decreases with $\beta$, which results from $a-b-c+d>0$, cf.\ \eq{weakUncond01R}. 
However, this is again consistent with the observation that  $\tau_1^A$ scales as $N \ln N$ in large populations. 
Although the fixation probability of a single $A$ decreases with $\beta$, if such an event occurs, it is faster than in the neutral case. 
A numerical example for this behavior is shown in Figure \fig{FIG2} (c).

The numerical examples indicate that the convergence radius of our weak selection expansion is of the 
order of $N^{-1}$, which is also known for many systems in population genetics. 
Although $N^{-1}$ might appear small, this kind of weak selection is the most relevant limit in evolutionary biology,
as evolutionary change is typically only connected with small selective differences.  
We stress that we have made no assumptions on the population size, such that our results are valid for
arbitrary $N$.

Our approach shows under which circumstances the general features of the game are reflected in the
fixation times under weak selection. Although the weak selection expansion of the mean exit or fixation times
is technically rather tedious, the resulting asymptotic behavior shows remarkable simplicity. 

\section*{Acknowledgment}
	Financial support by the Emmy-Noether program of the DFG is gratefully acknowledged.

\begin{figure}[B]
\begin{center}
	\includegraphics[height=0.9\textheight,angle=0]{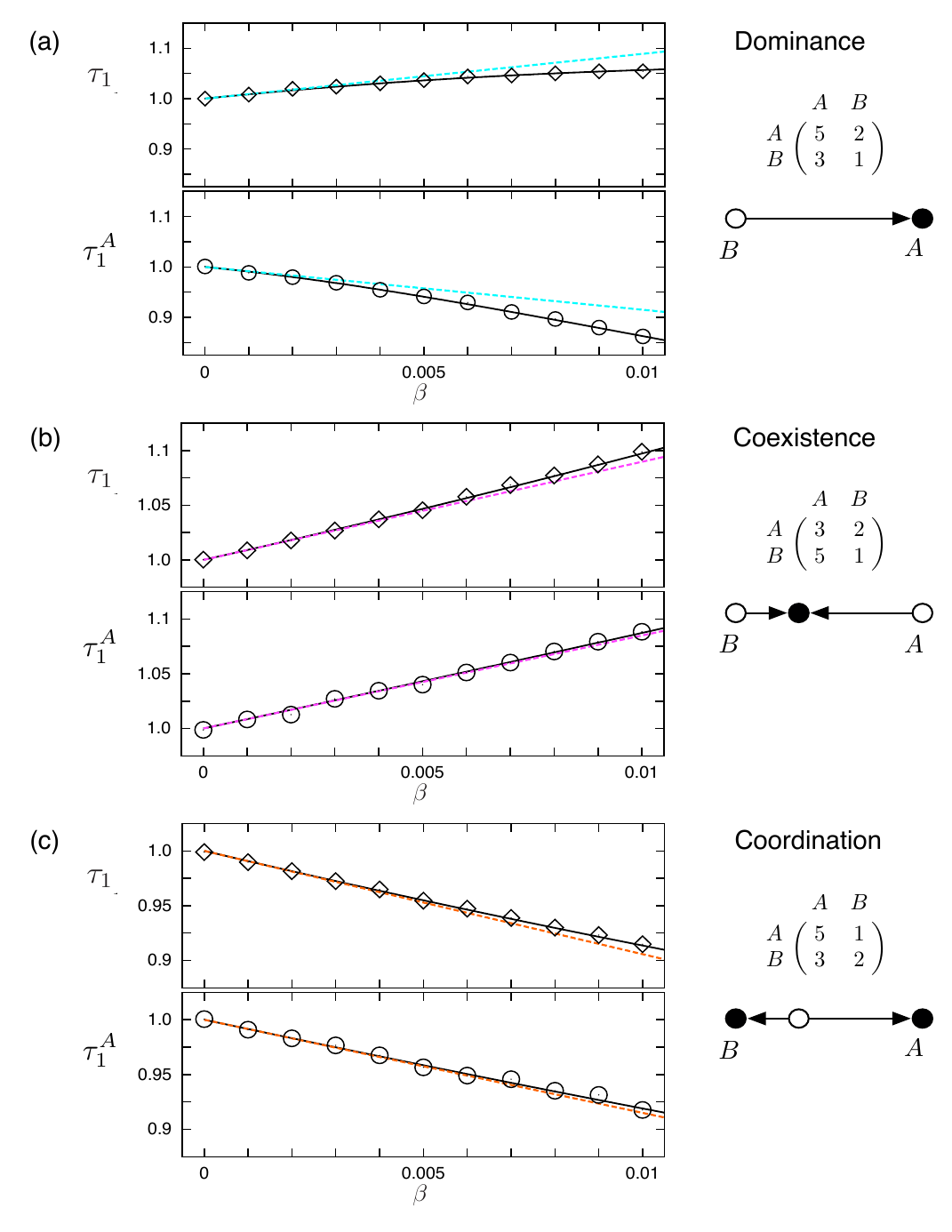}
	\end{center}
\caption{ 
Expectation values of the rescaled fixation times starting with a single $A$ mutant in a population of $B$ as a function of the selection intensity $\beta$.
Full lines show the normalized exact solution originating from the exact results \eq{weak01} and \eq{weak03a}.
Colored dashed lines are the linear approximations \eq{weakUncond01R} and \eq{weakCond01R}.
Symbols show the results from simulations based on $10^7$ realizations, which agree nicely with the exact results. Diamonds are for the unconditional averages, circles are for the conditional averages.
On the right hand side, we show the payoff matrices of the three games and illustrate the direction of selection in these games.
(a) In a game with dominance of strategy $A$, the unconditional fixation time increases with the intensity of selection, but the conditional fixation time decreases.
(b) For games with stable coexistence, both fixation times increase with the intensity of selection.
(c) For coordination games, the two fixation times become shorter when the intensity of selection is increased.
In all examples,
the population size is $N=100$.
}\label{fig:FIG2}
\end{figure}

\FloatBarrier

\begin{appendix}

\section{Finite double sums}\label{app:general}

Here, we collect some helpful calculations for double sums as they appear in the mean exit times.
An important observation is
\begin{eqnarray}\label{eq:Ptheorem}
	\sum\limits_{k=i}^{N-1}\sum\limits_{l=1}^k\frac{f_l}{N-l} = (N-i)\sum\limits_{l=1}^{i-1}\frac{f_l}{N-l}+\sum\limits_{l=i}^{N-1}f_l,
\end{eqnarray}
for any function $f_l<\infty$ and $l=1,\dots,N-1$. 
This can be seen by writing the left hand side term by term, \ie
\begin{eqnarray}\label{eq:Pthrmproof1}
	\sum\limits_{k=i}^{N-1}\sum\limits_{l=1}^k\frac{f_l}{N-l} &=& \frac{f_1}{N-1} + \dots + \frac{f_i}{N-i}\nonumber\\
	& + & \frac{f_1}{N-1}  + \dots+ \frac{f_i}{N-i} + \frac{f_{i+1}}{N-(i+1)}  \nonumber\\
	& + & \dots \nonumber \\
	& + &\frac{f_1}{N-1}  + \dots+ \frac{f_i}{N-i} +\dots+\frac{f_{N-1}}{N-(N-1)}\\
	&=&(N-i)\sum\limits_{l=1}^{i}\frac{f_l}{N-l}\nonumber\\
	&&+(N-i-1)\frac{f_{i+1}}{N-(i+1)} + \dots+f_{N-1}\nonumber\\
	&=&(N-i)\sum\limits_{l=1}^{i-1}\frac{f_l}{N-l}+\sum\limits_{l=i}^{N-1}f_l\nonumber.
\end{eqnarray}
For the case $i=1$ the result is especially simple, since the first sum of the right hand side of equation \eq{Ptheorem} vanishes. 
This case is of special interest for the computation of $t_1^A$ under neutral selection with $f_l=1$ and for $t_1$ with $f_l=1/l$.

Another finding for double sums with $M\in\mathbb N$ and two bounded functions $f_k$ and $g_l$ is
\begin{eqnarray}\label{eq:appsums01}
	\sum\limits_{k=1}^{M}\sum\limits_{l=1}^{k}f_k\,g_l=\sum\limits_{l=1}^{M}g_l\,\sum\limits_{k=l}^{M}f_k.
\end{eqnarray}
This becomes clear by resorting the terms again,
\begin{eqnarray}\label{eq:appsum01}
	\sum\limits_{k=1}^{M}\sum\limits_{l=1}^{k}f_k\,g_{l}&=&f_1g_1+f_2(g_1+g_2)
	+\dots+f_M(g_1+g_2+\dots+g_M)\nonumber\\
	&=&g_1(f_1+\dots+f_M)+g_2(f_2+\dots+f_M)+\dots+g_Mf_M\nonumber\\
	&=&\sum\limits_{l=1}^{M}g_l\,\sum\limits_{k=l}^{M}f_k.
\end{eqnarray}

\section{Fixation times under weak selection}

\label{app:weakA}

Here, we calculate the linear corrections of the mean exit times $t_1$ and $t_1^A$ for the Fermi process in detail, compare equations \eq{weak01} and \eq{weak03a}. We aim at finding these times for weak selection, e.g.\
\begin{eqnarray}
t_1 \approx  \left[t_1\right]_{\beta=0} + 
\beta
\left[\Del{}{\beta}\,t_1\right]_{\beta = 0} .
\end{eqnarray}
The first term follows directly from the calculation in \ref{app:general}, see equation \eq{t1N01}. 
Our goal here is to compute the linear term $\left[\Del{}{\beta}\,t_1\right]_{\beta = 0}$. 
\begin{eqnarray}\label{eq:appweak12}
	\left[\Del{}{\beta}\,t_1\right]_{\beta=0}&=& \sum\limits_{k=1}^{N-1}\sum\limits_{l=1}^{k}\left[ \frac{1}{T_l^+} \Del{ \phi_1}{\beta}+\phi_1\Del{}{\beta}\frac{1}{T_l^+}\right]_{\beta=0}\nonumber\\
	 &&-\sum\limits_{k=1}^{N-1}\sum\limits_{l=1}^{k}\left[\frac{\phi_1}{T_l^+}\sum\limits_{m=l+1}^{k}\Delta\pi(m)\right]_{\beta=0},
\end{eqnarray}
where we applied 
$\left[
\prod_{m=l+1}^{k}\gamma_m 
\right]_{\beta=0}=1$ 
and
$\left[ \Del{}{\beta}
\prod_{m=l+1}^{k}\gamma_m 
\right]_{\beta=0}=-\sum_{m=l+1}^{k}\Delta\pi(m)
$. 
For the fixation probability under weak selection and with $\Delta\pi(l)=u\,l+v$, we have 
\begin{eqnarray}\label{eq:appFixprobL01}
	\left[\Del{\phi_l}{\beta}\right]_{\beta=0}=\frac{l}{N}(N-l)\frac{(N+l)u+3v}{6}.
\end{eqnarray}
The weak selection approximation of the inverse of the transition probability $T_l^+$, compare equation \eq{TransProb}, yields
\begin{eqnarray}\label{eq:appWeakT}
	\left[ \Del{}{\beta}\frac{1}{T^+_l} \right]_{\beta=0}=-\frac{N^2}{l(N-l)}(u\, l+v).
\end{eqnarray}
Thisleads to
\begin{eqnarray}\label{eq:appWeakTX}
	\left[\Del{}{\beta}\,t_1\right]_{\beta=0}
	&=& \sum\limits_{k=1}^{N-1}\sum\limits_{l=1}^{k}\frac{(N-1)((N+1)u+3v)}{6N}\frac{2N^2}{l(N-l)}\nonumber\\
	&&-\sum\limits_{k=1}^{N-1}\sum\limits_{l=1}^{k}\frac{1}{N}\frac{N^2}{l(N-l)}\left( u\,l+v \right)\nonumber\\
	&&-\sum\limits_{k=1}^{N-1}\sum\limits_{l=1}^{k}\frac{1}{N}\frac{2N^2}{l(N-l)}\sum_{m=1+l}^{k}(u\,m+v)
\end{eqnarray}
While the first two double sums can be solved with the help of \ref{app:general}, the third term is
more complicated. For this more tedious calculation, we refer to \ref{app:weakB}.  
Eventually, the solution of the double and triple sums leads to
\begin{eqnarray}
	\left[\Del{}{\beta}\,t_1\right]_{\beta=0}
	&=&N(N-1)\frac{(N+1)u+3v}{3}H_{N-1}\nonumber\\
	&&-N(N-1)u-NH_{N-1}v\nonumber\\
	&&-N(N-1)\left(\frac{((N+1)u+3v)}{3}H_{N-1}-u-v \right)\nonumber\\
	&=&v\,N(N-1-H_{N-1}),
\end{eqnarray}
where the last step is elementary. Combining this with equation \eq{t1N01} leads finally to the unconditional mean exit time under weak selection, equation \eq{weakUncond01}.

For the conditional fixation time $t_1^A$, the linear term $\left[\frac{\del}{\del \beta}\,t_1^A\right]_{\beta=0}$ reads 
\begin{eqnarray}\label{eq:appweak13}
	\left[\frac{\del}{\del \beta}\,t_1^A\right]_{\beta=0}&=&\sum\limits_{k=1}^{N-1}\sum\limits_{l=1}^{k} \left[\frac{1}{T_l^+}\Del{ \phi_l}{\beta}+\phi_l\Del{}{\beta}\frac{1}{T_l^+}\right]_{\beta=0}\nonumber\\
	 &&-\sum\limits_{k=1}^{N-1}\sum\limits_{l=1}^{k} \left[\frac{\phi_l}{T_l^+}\sum\limits_{m=l+1}^{k}\Delta\pi(m)\right]_{\beta=0}.
\end{eqnarray}
The only difference compared to the unconditional fixation time, equation \eq{appweak12}, is the fixation probability $\phi_l$ instead of $\phi_1$. The linear term of the weak selection expansion of $\phi_l$ is given in equation \eq{appFixprobL01}. This yields
\begin{eqnarray}\label{eq:appWeakTY}
	\left[\frac{\del}{\del \beta}\,t_1^A\right]_{\beta=0}
	&=& \sum\limits_{k=1}^{N-1}\sum\limits_{l=1}^{k}\frac{l(N-l)((N+l)u+3v)}{6N}\frac{2N^2}{l(N-l)}\nonumber\\
	&&-\sum\limits_{k=1}^{N-1}\sum\limits_{l=1}^{k}\frac{l}{N}\frac{N^2}{l(N-l)}\left( u\,l+v \right)\nonumber\\
	&&-\sum\limits_{k=1}^{N-1}\sum\limits_{l=1}^{k}\frac{l}{N}\frac{2N^2}{l(N-l)}\sum_{m=l+1}^{k}(u\,m+v).
\end{eqnarray}
Again, the first two double sums can be solved using the results from \ref{app:general}.
The third term follows from a calculation which is similar to \ref{app:weakB}, but simpler. 
This last term reduces to
\begin{eqnarray}
\label{eq:b9}
&&\sum\limits_{k=1}^{N-1}\sum\limits_{l=1}^{k}\frac{l}{N}\frac{2N^2}{l(N-l)}\sum_{m=l+1}^{k}(u\,m+v)\nonumber\\
&&= N\frac{(N-2)(N-1)}{18}\left( (5N+3)u+9v \right).
\end{eqnarray}
Finally, combining the three terms again results in
\begin{eqnarray}
	\left[\frac{\del}{\del \beta}\,t_1^A\right]_{\beta=0}	&=&\frac{N^{2}(N-1)}{18}\left( (4N+1)u+9v \right)\nonumber\\
	&&-\frac{N(N-1)}{2}(N u+2v)\nonumber\\
	&&-(N-2)\frac{N(N-1)}{18}\left( (5N+3)u+9v \right)\nonumber\\
	&=&-u\,N(N-1)\frac{N^{2}+N-6}{18}.
\end{eqnarray}
In combination with equation \eq{t1AN01}, this results in the conditional mean exit time under weak selection, equation \eq{weakCond01}.

For completeness, we briefly repeat this calculation for the mean exit times of the frequency dependent Moran process. With the transition probabilities \eq{MoranT01} and \eq{MoranT02}, the fixation probabilities under weak selection are identical to those of the Fermi process, see equation \eq{FixprobL01}. However, the inverse transition probability is different in the weak selection regime, \ie the linear correction is
\begin{eqnarray}\label{eq:MoranTlin}
	\left[ \Del{}{\beta}\frac{1}{T^+_l} \right]_{\beta=0}=-\frac{N}{l}\Delta\pi(l)=-N\frac{u\,l+v}{l}.
\end{eqnarray}
Hence, for the unconditional mean exit time we have the same starting equation \eq{appweak12}. But with equation \eq{MoranTlin} this gives
\begin{eqnarray}\label{eq:appWeakTMoran01}
	\left[\Del{}{\beta}\,t_1\right]_{\beta=0}
	&=& \sum\limits_{k=1}^{N-1}\sum\limits_{l=1}^{k}\frac{(N-1)((N+1)u+3v)}{6N}\frac{N^2}{l(N-l)}\nonumber\\
	&&-\sum\limits_{k=1}^{N-1}\sum\limits_{l=1}^{k}\frac{1}{N}\,N\frac{u\,l+v}{l}\nonumber\\
	&&-\sum\limits_{k=1}^{N-1}\sum\limits_{l=1}^{k}\frac{1}{N}\frac{N^2}{l(N-l)}\sum_{m=1+l}^{k}(u\,m+v),
\end{eqnarray}
which differs from equation \eq{appWeakTX} only in the second double sum. With the previous findings for the Fermi processes times the required calculation is  straightforward and results in
\begin{eqnarray}\label{eq:appWeakTMoran02}
	\left[\Del{}{\beta}\,t_1\right]_{\beta=0}=v\,\frac{N}{2}\left( N+1-2H_{N} \right).
\end{eqnarray}
That is, this linear correction has a different dependence on the system size $N$.\\
For the conditional mean exit time the situation is similar. In difference to equation \eq{appWeakTY}, the linear correction reads
\begin{eqnarray}\label{eq:appWeakTMoran03}
	\left[\frac{\del}{\del \beta}\,t_1^A\right]_{\beta=0}
	&=& \sum\limits_{k=1}^{N-1}\sum\limits_{l=1}^{k}\frac{l(N-l)((N+l)u+3v)}{6N}\frac{N^2}{l(N-l)}\nonumber\\
	&&-\sum\limits_{k=1}^{N-1}\sum\limits_{l=1}^{k}\frac{l}{N}N\frac{u\,l+v}{l}\nonumber\\
	&&-\sum\limits_{k=1}^{N-1}\sum\limits_{l=1}^{k}\frac{l}{N}\frac{2N^2}{l(N-l)}\sum_{m=l+1}^{k}(u\,m+v).
\end{eqnarray}
This the leads to
\begin{eqnarray}
	\left[\frac{\del}{\del \beta}\,t_1^A\right]_{\beta=0}=-u\,\frac{N^{2}}{36}\left( N^{2}-3N+2 \right),
\end{eqnarray}
for the linear correction of the conditional mean exit times of the frequency dependent Moran process.

\section{Finite triple sum}
\label{app:weakB}

Here, we calculate the triple sum from \ref{app:weakA}, that require some additional steps. Our goal is to solve 
\begin{equation}
 \sigma=\sum\limits_{k=1}^{N-1}\sum\limits_{l=1}^{k}\frac{1}{l(N-l)} \sum_{m=1+l}^k\Delta\pi(m).
 \end{equation}
For the sum over payoff differences, we have
\begin{equation}
\sum_{m=1+l}^k\Delta\pi(m)=\sum_{m=1+l}^k(u\,m+v)=f_k-f_l,
\end{equation} 
where we introduced the function
\begin{eqnarray}\label{eq:appweak02}
	f_m=m(m+1)\frac{u}{2}+m\; v,
\end{eqnarray}
which is valid for any integer $m$.
Using partial fraction expansion, $ \frac{N}{l(N-l)} = \frac{1}{l} + \frac{1}{N-l}$, we obtain 
\begin{eqnarray}\label{eq:appweak03}
	 \sigma &=&
	 \sum\limits_{k=1}^{N-1}\sum\limits_{l=1}^{k}\frac{f_k}{l(N-l)}-\sum\limits_{k=1}^{N-1}\sum\limits_{l=1}^{k}\frac{f_l}{l(N-l)}\nonumber \\
	 &=&\frac{1}{N}  \underbrace{\sum\limits_{k=1}^{N-1}\sum\limits_{l=1}^{k}\frac{f_k}{l}}_{K_1}+\frac{1}{N}\underbrace{\sum\limits_{k=1}^{N-1}\sum\limits_{l=1}^{k}\frac{f_k}{N-l}}_{K_2}\nonumber\\
	 &&-\frac{1}{N}  \underbrace{\sum\limits_{k=1}^{N-1}\sum\limits_{l=1}^{k}\frac{f_l}{l}}_{K_3}-\frac{1}{N}\underbrace{\sum\limits_{k=1}^{N-1}\sum\limits_{l=1}^{k}\frac{f_l}{N-l}}_{K_4}.
\end{eqnarray}
We solve each part separately, starting with the last one. 
For $K_4$, we obtain with equation \eq{Ptheorem} from \ref{app:general}
\begin{eqnarray}\label{eq:appweak04}
	K_4 = \sum\limits_{k=1}^{N-1}\sum\limits_{l=1}^{k}\frac{f_l}{N-l}= \sum\limits_{k=1}^{N-1}f_k
	=\frac{N-1}{6}N((N+1)u+3v).
\end{eqnarray}
The second last term, $K_3$, is a sum over a linear function and can be treated with any table of elementary sums, \eg \cite{Graham94},
\begin{eqnarray}\label{eq:appweak05}
	K_3 &=& \sum\limits_{k=1}^{N-1}\sum\limits_{l=1}^{k}\frac{f_l}{l} = \sum\limits_{k=1}^{N-1}\sum\limits_{l=1}^{k}((l+1)\frac{u}{2}+v)\nonumber\\
	&=&\frac{N-1}{12}N((N+4)u+6v).
\end{eqnarray}
The remaining two terms require more effort. Both terms, $K_1$ and $K_2$ have the same structure regarding functions of $k$ and $l$. Using equation \eq{appsums01}, we have
\begin{eqnarray}\label{eq:appweak06}
	K_2 =\sum\limits_{k=1}^{N-1}\sum\limits_{l=1}^{k}\frac{f_k}{N-l}  =\,& \sum\limits_{l=1}^{N-1}\frac{1}{N-l}\,\sum\limits_{k=l}^{N-1}f_k,
\end{eqnarray}
and
\begin{eqnarray}\label{eq:appweak07}
	K_1= \sum\limits_{k=1}^{N-1}\sum\limits_{l=1}^{k}\frac{f_k}{l} =\,& \sum\limits_{l=1}^{N-1}\frac{1}{l}\,\sum\limits_{k=l}^{N-1}f_k.
\end{eqnarray}
Hence, we first have to compute the sum $\sum_{k=l}^{N-1}f_k$, which reduces to the solution of elementary sums, 
\begin{eqnarray}\label{eq:appwaek08}
	\sum\limits_{k=l}^{N-1}f_k &=\sum\limits_{k=l}^{N-1}k((k+1)\frac{u}{2}+v)
	=\frac{u}{2}\sum\limits_{k=l}^{N-1}k^2\,+\,\left( \frac{u}{2}+v \right)\sum\limits_{k=l}^{N-1}k\nonumber\\
	&=\frac{N-l}{6}\left (N^2+Nl+l^2-1)u+3(N+l-1)v \right)\nonumber\\
	&=\frac{N-l}{6}\left(N^2+Nl+l^2-1\right)u+\frac{N-l}{2}\left(N+l-1\right)v.
\end{eqnarray}
Thus, solving equations \eq{appweak06} and \eq{appweak07} simplifies to solving the elementary sums $\sum_{l=1}^{N-1}\,l^s$ with $s=0,1,2$, compare \cite{Graham94}. With this, we have
\begin{eqnarray}\label{eq:appweak09}
	K_{2} &=& \frac{N-1}{6}\sum\limits_{l=1}^{N-1} \left((N+1)u+3v \right)+\frac{Nu+3v}{6}\sum\limits_{l=1}^{N-1}\,l+\frac{u}{6}\sum\limits_{l=1}^{N-1}l^2\nonumber\\
	&=&\frac{N-1}{36}\left( (11N^2-N-6)u+9(3N-2)v \right).
\end{eqnarray}
For $K_1$, we obtain
\begin{eqnarray}\label{eq:appweak10}
	K_1&=&\frac{1}{6}\sum\limits_{l=1}^{N-1} \frac{N(N-1)(N+1)u+3N(N-1)v}{l}\nonumber\\
	&&-\frac{1}{6}\sum\limits_{l=1}^{N-1}\left((l^2-1)u - 3(l-1)v\right)\nonumber\\
	&=&\frac{N(N-1)}{6}\left((N+1)u+3v\right)\,H_{N-1}\nonumber\\
	&&-\frac{N-1}{36}(N-2)((2N+3)u+9v). \nonumber
\end{eqnarray}
Summing up the terms, $\sigma=(K_1+K_2-K_3-K_4)/N$, finally yields the result
\begin{eqnarray}\label{eq:appweak03a}
	 \sigma &=&\frac{N-1}{6}\left( ((N+1)u+3v)H_{N-1}-3(u+v) \right).
\end{eqnarray}
Again,  $H_n=\sum_{l=1}^{n}1/l$ are the harmonic numbers. 
In equation \eq{b9}, the reasoning is very similar, but only terms of the structure of $K_2$ and $K_4$ appear.

\end{appendix}

\end{document}